\documentclass{eptcs}
\providecommand{\event}{TFPIE} 

\usepackage{amsmath}

\newcommand{\abs}[1]{\left| #1 \right|}

\title{Learn Physics by Programming in Haskell}
\author{Scott N. Walck
\institute{Department of Physics\\
  Lebanon Valley College\\
  Annville, Pennsylvania, USA}
\email{walck@lvc.edu}
}

\def\titlerunning{Learn Physics by Programming in Haskell}
\def\authorrunning{Scott N. Walck}

\begin{document}

\maketitle

\begin{abstract}

We describe a method for deepening a student's understanding
of basic physics by asking the student to express physical ideas
in a functional programming language.
The method is implemented in a second-year course in computational
physics at Lebanon Valley College.
We argue that the structure of Newtonian mechanics is clarified by its expression in
a language (Haskell) that supports higher-order functions, types, and
type classes.
In electromagnetic theory, the type signatures of functions that calculate
electric and magnetic fields clearly express the functional dependency
on the charge and current distributions that produce the fields.
Many of the ideas in basic physics are well-captured by a type
or a function.

\end{abstract}

\section{Introduction}

Introduction to Computational Physics (Physics 261) at Lebanon Valley College
is a second-year one-semester elective physics course.  The prerequisites for the course
are one year of introductory physics and one semester of calculus.
\emph{No previous programming experience is expected or assumed.}
Nevertheless, third- and fourth-year students often take the course, and
a number of students do have previous programming experience (although
not in a functional language).

The purpose of the course is to strengthen a student's understanding of basic
physics by learning a new language (Haskell), and instructing the computer
to do physics in that language.
Our attitude was strongly influenced by the work of Papert\cite{papert},
and the subsequent work of Sussman and his coworkers\cite{sicm,sussmanFDG}.
A functional programming language is a good choice for this purpose for a number
of reasons.  Functional programming languages tend to look, or at least feel,
a bit more like mathematics than imperative languages.  The concept
of a list is quickly and comfortably assimilated, and standard library list
functions take the place of loops in an imperative language.
Referential transparency removes much of the convolution
of thinking
about code.  Many of the concepts that appear in introductory physics
have a natural expression as types or higher-order functions.

Haskell is a particularly good choice because its surface syntax appears to
be familiar mathematics (compared with Scheme, say), and its system
of static types aids the programmer in thinking about the structure
of what she is writing.  Haskell's system of curried functions is
very pleasant and convenient.
Type classes are not essential for the purposes
of Physics 261, but they come in handy in a few places.

The course is structured in roughly two parts.  In the first part, we learn a subset
of the Haskell programming language.
We are particularly interested in types, functions, and higher-order functions.
We introduce a number of the types and functions provided by the standard Prelude,
and we focus on how to write our own functions.
In previous offerings of the course, we used the first five chapters
of Hutton's book \cite{hutton07}.
We are less interested in type classes, but we need to be aware of them
to understand the types of some functions, and to have any chance of understanding
error messages.  We intentionally avoid explicit recursion because we're
only spending about seven weeks learning our subset of Haskell.
We also omit the creation of new algebraic data types.
Nevertheless, most students seem able to become proficient in this subset
of Haskell in about seven weeks, and can then apply it to problems
in mechanics and electromagnetic theory.

In the second part of the course, we use Haskell
to express the ideas of Newtonian mechanics and electromagnetic theory.
Here we want students to use the language as a set of building blocks
for constructing interesting things, and we want to provide a lot of freedom
for students to use the language as they see fit.
At the same time, we have in mind
a way of viewing Newtonian mechanics and (part of) electromagnetic theory
toward which we are guiding students.
Section 2 describes our view of Newtonian mechanics, and shows
what functional programming has to offer toward it.
Section 3 describes ways in which types and higher-order functions
serve to organize and clarify parts of electromagnetic theory.
Section 4 is a short conclusion.

\section{Newtonian Mechanics}

Newton's second law, $F_{\rm net} = m a$, appears deceptively simple.
Coupled with the natural tendency in a first physics course
to focus on problems that are easily and analytically solvable,
the structure of Newtonian mechanics usually gets missed.

We take a state-based approach to Newtonian mechanics.
The state of a single particle is given by its position and its
velocity (or, equivalently, its momentum).  States of rigid bodies
(able to rotate) and of multiple particles or bodies require more information.
In any case, the state is characterized by a \emph{type};
initial conditions for the problem are specified by a value of that type.

\subsection{Vectors}

Three-dimensional vectors play a central role in Newtonian mechanics.
They are used to describe velocity, acceleration, force, and momentum.

We define a data type for three-dimensional vectors.
\footnote{Most of the code presented in this paper can be found in the
learn-physics package\cite{learn-physics}.}
\begin{verbatim}
data Vec = Vec { xComp :: Double
               , yComp :: Double
               , zComp :: Double }
\end{verbatim}

In the Physics 261 course, we introduce vectors in two stages.
In stage 1, we introduce functions that apply only to
the \verb|Vec| data type, giving these operations easily digestible
type signatures that clearly express their purpose.  These type
signatures are shown in Table \ref{vectorfunctions}.
Some of the operators in Table \ref{vectorfunctions}
are provided for convenience; for example
we provide a scalar multiplication operator \verb|(*^)| in which
the scalar goes on the left and the vector on the right,
as well as an alternative version \verb|(^*)| in which the
arguments are flipped.  This redundancy matches the operators
in Conal Elliott's vector-space package\cite{vector-space}.

\begin{table}
\begin{center}
\begin{tabular}{lll}
Function      & Description           & Type \\ \hline
\verb|(^+^)| & vector addition       & \verb|Vec -> Vec -> Vec| \\
\verb|(^-^)| & vector subtraction    & \verb|Vec -> Vec -> Vec| \\
\verb|(*^)|  & scalar multiplication & \verb|Double -> Vec -> Vec| \\
\verb|(^*)|  & scalar multiplication & \verb|Vec -> Double -> Vec| \\
\verb|(^/)|  & scalar division       & \verb|Vec -> Double -> Vec| \\
\verb|(<.>)| & dot product           & \verb|Vec -> Vec -> Double| \\
\verb|(><)|  & cross product         & \verb|Vec -> Vec -> Vec| \\
\verb|magnitude| & magnitude        & \verb|Vec -> Double| \\
\verb|zeroV| & zero vector          & \verb|Vec| \\
\verb|iHat|  & unit vector           & \verb|Vec| \\
\verb|negateV| & vector negation    & \verb|Vec -> Vec| \\
\verb|vec|  & vector construction    & \verb|Double -> Double -> Double -> Vec| \\
\verb|xComp|  & vector component    & \verb|Vec -> Double| \\
\verb|sumV|   & vector sum          & \verb|[Vec] -> Vec| \\
\end{tabular}
\end{center}
\caption{Functions for working with vectors, stage 1.
In stage 1, all of the functions for working with vectors have concrete types.
This makes their type signatures easier for students to read and reason about.
}
\label{vectorfunctions}
\end{table}

Later in the course, around the time we want to write a numerical integrator that can
work with both scalars and vectors, we introduce stage 2 in our description of
vectors.  In stage 2, we keep the \verb|Vec| data type, but redefine the functions
that act on it to belong to type classes from the vector-space package\cite{vector-space}
that can accommodate numbers as well as
\verb|Vec|s.  The cost of this abstraction is that the type signatures of these functions are now 
more difficult to read, and require an understanding of type classes.
The stage 2 type signatures are shown in Table \ref{vectorfunctionsstage2}.

\begin{table}
\begin{center}
\begin{tabular}{lll}
Function      & Description           & Type \\ \hline
\verb|(^+^)| & vector addition       & \verb|AdditiveGroup v => v -> v -> v| \\
\verb|(^-^)| & vector subtraction    & \verb|AdditiveGroup v => v -> v -> v| \\
\verb|(*^)|  & scalar multiplication & \verb|VectorSpace v => Scalar v -> v -> v| \\
\verb|(^*)|  & scalar multiplication & \verb|VectorSpace v => v -> Scalar v -> v| \\
\verb|(^/)|  & scalar division       & \verb|(VectorSpace v, Fractional (Scalar v))| \\
              &                      & \verb| => v -> Scalar v -> v| \\
\verb|(<.>)| & dot product           & \verb|InnerSpace v => v -> v -> Scalar v| \\
\verb|magnitude| & magnitude        & \verb|(InnerSpace v, Floating (Scalar v))| \\
              &                      & \verb| => v -> Scalar v| \\
\verb|zeroV| & zero vector          & \verb|AdditiveGroup v => v| \\
\verb|negateV| & vector negation    & \verb|AdditiveGroup v => v -> v| \\
\verb|sumV|   & vector sum          & \verb|(Foldable f, AdditiveGroup v)| \\
               &                     & \verb| => f v -> v| \\
\end{tabular}
\end{center}
\caption{Functions for working with vectors, stage 2.
In stage 2, we want to be able to write code, such as a numerical integrator,
that can work with numbers or vectors.
The type classes are defined in Conal Elliott's vector-space package\cite{vector-space}.
}
\label{vectorfunctionsstage2}
\end{table}

\subsection{Single-particle mechanics}

The state of a single particle in three dimensions can be specified by giving
the particle's position and the particle's velocity.
It is convenient to include the current time as part of the state as well;
this allows the inclusion of time-dependent forces that may act
on the particle.  It is also convenient to work with the displacement
from some chosen origin, rather than working directly with position,
because position is not a vector (it makes no sense to add positions),
but displacement is a vector.  The state of our system of one particle
can then be expressed as follows.

\begin{verbatim}
type Time         = Double
type Displacement = Vec
type Velocity     = Vec
type State        = (Time, Displacement, Velocity)
\end{verbatim}

The state of a particle changes based on the local forces that act on it.
We would like to express this idea with a function
\verb|Double -> State -> State| that takes a short time interval
and updates the state accordingly.

What information is needed to know how to update the state?
We need to know the forces that act on the particle; from the
net force on the particle and the mass of the particle, Newton's
second law allows us to calculate the acceleration of the particle.
The key information is contained in what we call an
\verb|AccelerationFunction|.

\begin{verbatim}
type AccelerationFunction = State -> Vec
\end{verbatim}

With an \verb|AccelerationFunction|, we have specified a
system of first-order differential equations; the rate of change
of displacement is given by velocity, and the rate of change of
velocity is given by acceleration.
Using the Euler method to solve the differential equation,
we arrive at the following function.  (We start with the Euler
method because it is the simplest and most intuitive to understand.)

\begin{verbatim}
eulerStep :: AccelerationFunction -> Double -> State -> State
eulerStep a dt (t,r,v) = (t',r',v')
    where
      t' = t + dt
      r' = r ^+^ v ^* dt
      v' = v ^+^ a(t,r,v) ^* dt
\end{verbatim}

To define any particular one-particle problem, we have only to specify the appropriate
acceleration function.
For a satellite orbiting a fixed Earth, for example, we have
the following function to produce the satellite's acceleration from
the current state of the satellite.
\begin{verbatim}
satellite :: AccelerationFunction
satellite (t,r,v) = 6.67e-11 * 5.98e24 / magnitude r ^ 2 *^ u
    where
      u = negateV r ^/ magnitude r
\end{verbatim}
Here, the universal gravitational constant and the mass of Earth
are expressed numerically in SI units.  We see the inverse square law
for universal gravity.  The unit vector \verb|u|
points from the satellite toward the Earth.
(The negation is because the displacement vector \verb|r| points
from the Earth to the satellite.)  Notice that while the state
consists of time, displacement, and velocity, the force and acceleration
in this problem depend only on displacement, and not on time or velocity.

Another one-particle problem is the damped, driven, harmonic oscillator.
The particle in this situation is subject to three forces---a spring force,
a damping force, and a driving force.
An acceleration function for this situation could be written as follows.
\begin{verbatim}
dampedDrivenOsc :: Double  -- damping constant
                -> Double  -- drive amplitude
                -> Double  -- drive frequency
                -> AccelerationFunction
dampedDrivenOsc beta driveAmp omega (t,r,v)
    = (forceDamp ^+^ forceDrive ^+^ forceSpring) ^/ mass
      where
        forceDamp   = (-beta) *^ v
        forceDrive  = driveAmp * cos (omega * t) *^ iHat
        forceSpring = (-k) *^ r
        mass        = 1
        k           = 1  -- spring constant
\end{verbatim}
Here we have decided to commit to numeric values for some of the parameters, such
as the mass and spring constant, while passing others as parameters
to the function \verb|dampedDrivenOsc|.
Note that the net force for the damped, driven, harmonic oscillator
depends on time, displacement, and velocity.  The driving force is
in the $x$~direction.

We can obtain a solution as an infinite list of \verb|State| values by iterating
the \verb|eulerStep| function we defined above.  Shown here is a related function
\verb|eulerCromerStep| which uses the Euler-Cromer method
(see, for example \cite{giordano97}), an improved
version of the Euler method.

\begin{verbatim}
solution :: AccelerationFunction -> Double -> State -> [State]
solution a dt = iterate (eulerCromerStep a dt)

states :: [State]
states = solution (dampedDrivenOsc 0 1 0.7) 0.01
         (0, vec 1 0 0, vec 0 0 0)
\end{verbatim}

Once in possession of a list of \verb|State|s,
we can pick out relevant data for plotting or animation.
For example, here is a function to form a list of pairs of times and $x$~components
of displacement, to make a plot of $x$ vs. $t$.
\begin{verbatim}
txPairs :: [State] -> [(Double,Double)]
txPairs sts = [(t, xComp r) | (t,r,v) <- sts]
\end{verbatim}

Any single-particle problem in three dimensions can be treated in a similar
way.  The specification of a particular situation consists in writing
an \verb|AccelerationFunction|, as we did above for
satellite motion and for the damped, driven, harmonic oscillator.

\subsection{Beyond a single-particle state space}

We want to study problems in mechanics that go beyond
a single particle in three dimensions.
For such problems, our first task is to decide on an appropriate
type to characterize the state of our system.
For a system of multiple particles, one choice is to pair
the current time with a list of displacement-velocity pairs
(one pair for each particle).
\begin{verbatim}
type SystemState = (Time, [(Displacement, Velocity)])
\end{verbatim}
This type has the advantage that it works for any number of particles
(and the disadvantage that Haskell's type system will not warn us
if our code erroneously changes the length of the particle list).
For the \verb|SystemState| above, the crucial information needed
to update the state is a list of particle accelerations.
Taking the place of \verb|AccelerationFunction|, then, is
\verb|SystemAccFunc|, which provides an acceleration for each of
the particles as a function of the system state.
We also generalize our \verb|eulerStep| or \verb|eulerCromerStep| numerical method
to handle the new system state.
\begin{verbatim}
type SystemAccFunc = SystemState -> [Vec]

eulerCromerSystemStep :: SystemAccFunc -> Double -> SystemState -> SystemState
eulerCromerSystemStep a dt (t,rvs) = (t + dt,rvs')
    where
      as = a (t,rvs)
      (rs,vs) = unzip rvs
      rs' = zipWith (^+^) rs (map (^* dt) vs')
      vs' = zipWith (^+^) vs (map (^* dt) as)
      rvs' = zip rs' vs'
\end{verbatim}

Using techniques like this, we simulate in the course
a system consisting of the Sun, Earth, and Moon mutually interacting
by universal gravitation.
Increasing the number of particles to about 100, we also model
an elastic vibration as a collection of point masses with nearest-neighbor
Hooke's-law spring interactions.\cite{giancoliPSE4p318,goldstein12p1}
A physical pendulum is another problem we study, where the state is best
expressed with an angle and an angular velocity for the pendulum in
its rotation about a fixed pivot.

\subsection{Uncoupling the numerical solution method}

There are many numerical methods that one can use to solve a differential
equation.  In the exposition above, we have applied the Euler method
(or the Euler-Cromer method) in the same function that effectively sets
up the differential equation to be solved.  This has some advantage
for students who have not yet studied differential equations; the Euler
method is particularly easy to read and understand.

Nevertheless, it would be conceptually cleaner to separate the construction of the differential
equation from its solution with a particular numerical method.
Moreover, there is nothing approximate in the construction of the differential
equation from the forces that are present in a given physical situation,
while numerical solution methods are invariably approximate.
In fact, a clean separation can be made.  Although
we have never had time to address this issue in the course,
it is a desirable way to organize our thinking, and extends the
theme of exposing the structure of Newtonian mechanics.

We can define a type class \verb|StateSpace|
for data types that can serve as the state of a physical system.
This type class is a bit more general than the \verb|VectorSpace| class,
because we want data types for position, which is not a vector,
to be able to be part of the state.
If \verb|state| is an instance of \verb|StateSpace|,
there is an associated data type \verb|Diff state| that represents
the vector space of time derivatives of the state space.
The \verb|StateSpace| type class is a modification of the
\verb|AffineSpace| type class\cite{vector-space}, in which the
scalars of the associated vector space are required to be instances
of \verb|Fractional|.

A differential equation is then simply a function from the state space to
its linearization giving the time derivatives of each of the dependent
variables in the state.
\begin{verbatim}
type DifferentialEquation state = state -> Diff state
\end{verbatim}
For example, for the single-particle \verb|State| that we defined earlier,
we must specify the time derivatives of time, displacement, and velocity.
This is easily done in terms of our \verb|AccelerationFunction|.
\begin{verbatim}
oneParticleDiffEq :: AccelerationFunction -> DifferentialEquation State
oneParticleDiffEq a (t, r, v) = (1, v, a(t, r, v))
\end{verbatim}

An evolution method is a way of approximating the state
after advancing a finite interval in the independent
variable (time) from a given state.
\begin{verbatim}
type EvolutionMethod state
    = DifferentialEquation state  -- ^ differential equation
    -> Scalar (Diff state)        -- ^ time interval
    -> state                      -- ^ initial state
    -> state                      -- ^ evolved state
\end{verbatim}
The Euler method is the simplest evolution method.
\begin{verbatim}
eulerMethod :: StateSpace state => EvolutionMethod state
eulerMethod de dt st = st .+^ de st ^* dt
\end{verbatim}
Here the operator \verb|(.+^)| is borrowed from \verb|AffineSpace|
to represent shifting a point in the state space by a vector from
the associated vector space.
Other evolution methods, such as a Runge-Kutta method, can
now be easily plugged in.

To obtain a full solution, we define two more type synonyms,
and a function to form an iterative solution.
\begin{verbatim}
type InitialValueProblem state = (DifferentialEquation state, state)
type SolutionMethod state = InitialValueProblem state -> [state]

stepSolution :: EvolutionMethod state
             -> Scalar (Diff state)    -- time interval
             -> SolutionMethod state
stepSolution ev dt (de, ic) = iterate (ev de dt) ic
\end{verbatim}
An initial value problem is a differential equation along with an initial state.
A (numerical) solution method is a way of converting
an initial value problem into a list of states (a solution).
The infinite list \verb|states| of state values that we obtained above would now be formed
as follows.
\begin{verbatim}
states' :: [State]
states' = stepSolution eulerMethod 0.01
          ( oneParticleDiffEq (dampedDrivenOsc 0 1 0.7)
          , (0, vec 1 0 0, vec 0 0 0) )
\end{verbatim}

\subsection{Mechanics summary}

We summarize our state space view of mechanics by giving a three-step
process for analyzing a physical situation.

\begin{enumerate}
\item Choose a type to represent the state space for the problem.
This involves a choice of which quantities to pay attention to,
but it does not require a knowledge of the nature of the forces
that are acting.
\item Describe how the state changes in time.  Newton's second law
(and possibly also Newton's third law) is at the heart of this description.
\item Give an initial state for the system.
Now we can make graphs or animations.
\end{enumerate}

\section{Electromagnetic Theory}

One type of problem that occurs in electromagnetic theory is the
calculation of electric and magnetic fields produced by charge
and current distributions.

\subsection{Electric field produced by a continuous charge distribution}

The electric field at position $\vec{r}$ produced by a charge distributed over a curve $C$ is
\begin{equation}
\vec{E}(\vec{r}) = \frac{1}{4 \pi \epsilon_0} \int_C \lambda(\vec{r}') \frac{\vec{r} - \vec{r}'}{\abs{\vec{r} - \vec{r}'}^3} dl' ,
\label{efield}
\end{equation}
where $\lambda(\vec{r}')$ is the linear charge density at $\vec{r}'$
and $\epsilon_0$ is a constant called the permittivity of free space.

Unlike in mechanics, where the simple expression of Newton's second law hides
the complexity of the situation, here the notation is daunting.
One issue in particular that the notation hides from the beginning student
is that in order to calculate the electric field, we need to have (1)
a curve $C$ describing where the charge is, and (2) a charge density $\lambda$,
and \emph{that is all}.

Since curves are important in a number of places in electromagnetic theory,
we define a data type for curves.
\begin{verbatim}
data Curve = Curve { curveFunc          :: Double -> Position
                   , startingCurveParam :: Double
                   , endingCurveParam   :: Double }
\end{verbatim}
A \verb|Curve| is a parametrized curve along with starting and ending
parameters.  The parametrized curve is expressed here in terms of a
\verb|Position| data type, defined similarly to \verb|Vec|, but without
the vector operations.

Here are some example curves.
\begin{verbatim}
circularLoop :: Double -> Curve
circularLoop radius
    = Curve (\t -> cart (radius * cos t)
                        (radius * sin t)
                         0
            ) 0 (2*pi)

line :: Double -> Curve
line l = Curve (\t -> cart 0 0 t) (-l/2) (l/2)
\end{verbatim}
The function \verb|cart| produces a \verb|Position| by giving Cartesian coordinates.

Equipped with our \verb|Curve| data type, we next need a way to integrate over a curve.
A line integral can have either a scalar or vector integrand
(to calculate electric field, we'll use a vector integrand, but to
calculate electric potential, we would use a scalar integrand).
\begin{align*}
\int_C f(\vec{r}') dl' \hspace{10mm} & \textrm{or} \hspace{10mm} \int_C \vec{F}(\vec{r}') dl'
\end{align*}

We use type synonyms to describe what a scalar field is and
what a vector field is.
\begin{verbatim}
type ScalarField = Position -> Double
type VectorField = Position -> Vec
type Field v     = Position -> v
\end{verbatim}
A scalar field is an assignment of a scalar (number) to each position in space.
A vector field is an assignment of a vector to each position in space.
We also want to be able to refer to a field that could be either
a scalar field or a vector field, so we define \verb|Field v|.

I usually ask students to write a simple numerical integrator
using the trapezoidal rule, to get a sense for what is required.
Then, I ask them to use code that I have written to integrate over a curve.
We use a general purpose line integral with the following type.

\begin{verbatim}
simpleLineIntegral
    :: (InnerSpace v, Scalar v ~ Double)
       => Int      -- ^ number of intervals
    -> Field v     -- ^ scalar or vector field
    -> Curve       -- ^ curve to integrate over
    -> v           -- ^ scalar or vector result
\end{verbatim}
The integrator works by chopping the curve into
a number to intervals, evaluating the field on each interval,
multiplying each value by the length of its interval, and summing.

Now we can calculate the electric field (\ref{efield}) of a one-dimensional charge distribution.
\begin{verbatim}
eFieldFromLineCharge
    :: ScalarField     -- ^ linear charge density lambda
    -> Curve           -- ^ geometry of the line charge
    -> VectorField     -- ^ electric field (in V/m)
eFieldFromLineCharge lambda c r
    = k *^ simpleLineIntegral 1000 integrand c
      where
        k = 9e9  -- 1 / (4 * pi * epsilon0)
        integrand r' = lambda r' *^ d ^/ magnitude d ** 3
            where
              d = displacement r' r
\end{verbatim}

Notice how types and higher-order functions are an essential aspect of
this definition.  The types \verb|ScalarField| and \verb|VectorField|
are themselves functions, one used as input and one used as output
of the \verb|eFieldFromLineCharge| function.  The \verb|Curve|
type, like \verb|Vec| and \verb|Position|, is used to package
information that, from the perspective of physics, rightfully
belongs together.
Also note how the type signature describes that the charge density
$\lambda$ and the curve $C$ are the only inputs necessary to make the
calculation.

\subsection{Magnetic field produced by a current-carrying wire}

Let us take the example of the magnetic field produced by
a current-carrying wire.  We will allow the wire to be of any
shape, and to carry any amount of current.  We wish to calculate
the magnetic field produced by the wire.
The magnetic field at position $\vec{r}$ produced by a current $I$ flowing along a curve $C$ is
given by the Biot-Savart law.
\begin{equation}
\vec{B}(\vec{r}) = \frac{\mu_0 I}{4 \pi} \int_C \frac{d\vec{l}' \times (\vec{r} - \vec{r}')}{\abs{\vec{r} - \vec{r}'}^3}
\label{biotsavart}
\end{equation}

To implement the Biot-Savart law, it is useful to have a
general purpose ``crossed line integral''.
\[
\int_C \vec{F}(\vec{r}') \times d\vec{l}'
\]
\begin{verbatim}
-- | Calculates integral vf x dl over curve.
crossedLineIntegral
    :: Int          -- ^ number of intervals
    -> VectorField  -- ^ vector field
    -> Curve        -- ^ curve to integrate over
    -> Vec          -- ^ vector result
\end{verbatim}

Do we really need a new integrator for the crossed line integral?
Can we not write 
\[
\int_C \vec{F}(\vec{r}') \times d\vec{l}' = \int_C \vec{F}(\vec{r}') \times \hat{t} dl'
\]
where $\hat{t}$ is a unit tangent to the curve, and use our previous integrator,
applying the cross product in the integrand?
That is a possibility, but I did not want to burden the user with supplying
tangent vectors along the curve.  Part of the motivation for the functions
which calculate electric and magnetic fields is to show \emph{how little}
is required to make the calculation.

Here is an implementation of the Biot-Savart law (\ref{biotsavart}).
\begin{verbatim}
bFieldFromLineCurrent
    :: Current      -- ^ current (in Amps)
    -> Curve        -- ^ geometry of the line current
    -> VectorField  -- ^ magnetic field (in Tesla)
bFieldFromLineCurrent i c r
    = k *^ crossedLineIntegral 1000 integrand c
      where
        k = 1e-7  -- mu0 / (4 * pi)
        integrand r' = (-i) *^ d ^/ magnitude d ** 3
            where
              d = displacement r' r
\end{verbatim}
Here, \verb|Current| is a synonym for \verb|Double|.
\begin{verbatim}
type Current = Double
\end{verbatim}
The minus sign prefixing the current in the integrand is because the crossed line integral
performs the cross product in the opposite order from the Biot-Savart law.
The magnetic field produced by a circular current loop is a great problem for
numerical investigation; despite the symmetry, the magnetic field from a circular loop is not
analytically calculable (except on the axis of symmetry).
Finally, note how the type signature shows that a current $I$
and a curve $C$ are all the information needed to calculate magnetic field.

\section{Conclusion and Future Work}

We have shown some of the ways that we have used Haskell to deepen a student's
understanding of Newtonian mechanics and parts of electromagnetic theory.
Types and higher-order functions have been essential in this method,
and have been used to describe ideas such as state spaces, curves,
vector fields, and methods for calculation.

One obvious use of types in physics that we have not explored in this work
is the expression of physical dimensions (length, mass, time) and units
(meter, kilogram, second).  Allowing the expression of units is very
desirable from a pedagogical perspective.  This is not trivial to do
with Haskell's type system because one wants multiplication to
``multiply the units'' as well as the numbers.  Nevertheless, there
are some Haskell libraries available and being developed for this purpose,
and we intend to explore their suitability to complement the ideas
in this paper.

\bibliographystyle{eptcs}
\bibliography{lppfullnomint}

\end{document}